\documentclass[aps,prl,reprint,superscriptaddress,nofootinbib,preprintnumbers]{revtex4-2}

\usepackage{amsmath}
\usepackage{amssymb}
\usepackage{booktabs}
\usepackage{braket}
\usepackage{dcolumn}
\usepackage{graphicx}
\usepackage[utf8]{inputenc}
\usepackage{siunitx}
\usepackage{xcolor}
\usepackage{xspace}
\usepackage{hyperref}
\usepackage{multirow}
\usepackage[nameinlink,capitalize]{cleveref}

\newcommand{\eg}{\textit{e.g.}\xspace}
\newcommand{\ie}{\textit{i.e.}\xspace}
\newcommand{\EOS}{\texttt{EOS}\xspace}

\newcommand{\Fpi}{\ensuremath{F_\pi^{I=1}}}
\newcommand{\GeV}{\text{GeV}}
\newcommand{\MeV}{\text{MeV}}
\renewcommand{\Re}{\operatorname{Re}}
\renewcommand{\Im}{\operatorname{Im}}

\begin{document}

\title{A Simple Parametrisation of the Pion Form Factor}
\author{Matthew Kirk}
\email{matthew.j.kirk@durham.ac.uk}
\affiliation{Institute for Particle Physics Phenomenology and Department of Physics, Durham University, Durham DH1 3LE, UK}
\author{Bastian Kubis}
\email{kubis@hiskp.uni-bonn.de}
\affiliation{Helmholtz-Institut für Strahlen- und Kernphysik (Theorie) and\\ Bethe Center for Theoretical Physics, Universität Bonn, 53115 Bonn, Germany}
\author{M\'eril Reboud}
\email{merilreboud@gmail.com}
\affiliation{Université Paris-Saclay, CNRS/IN2P3, IJCLab, 91405 Orsay, France}
\author{Danny van Dyk}
\email{danny.van.dyk@gmail.com}
\affiliation{Institute for Particle Physics Phenomenology and Department of Physics, Durham University, Durham DH1 3LE, UK}

\begin{abstract}
    We discuss a novel and simple parametrisation of the pion vector form factor that transparently connects
    spacelike and timelike regions of the momentum transfer $q^2$.
    Our parametrisation employs the framework of conformal mapping and respects the known analyticity properties of the
    form factor, accounting explicitly for the $\rho(770)$-meson pole.
    The parametrisation manifestly fulfils the normalisation condition at $q^2 = 0$ as well as further
    restrictions at the pion production threshold and in the limit $|q^2| \to \infty$.
    In contrast to the widely used Omnès parametrisation, our approach does not use the
    pion--pion scattering phase shift as input.
    We confront the parametrisation with experimental data from $\pi H$ scattering and
    $\tau^- \to \pi^-\pi^0\nu$ decay. We already find a good description of the data 
    with only five free parameters, which include the pole mass and decay width of the $\rho(770)$.
\end{abstract}

\preprint{%
    EOS-2024-03, IPPP/24/65
}

\maketitle

\section{Preliminaries}
\label{sec:prelim}

Hadronic form factors are scalar-valued functions that emerge
in a variety of processes such as $e^+e^- \to \pi\pi$, or $\bar{B}\to \pi \ell^-\bar{\nu}$.
They parametrise the mismatch between the partonic amplitude and the hadronic one,
cannot be computed perturbatively, and commonly contribute substantially to the theory uncertainties in such processes.

We set out to use a series expansion approach to parametrise a hadronic form factor
both below and above its respective pair production threshold,
whilst fulfilling a dispersive bound.
A past attempt at this idea was instructive but ultimately unsuccessful~\cite{Buck:1998kp}.
The salient difference between this past approach and our proposed parametrisation is our explicitly accounting
for above-threshold poles on the 2nd Riemann sheet. In this, we make use of the analytic structure of the hadronic form factors
as discussed in Ref.~\cite{Caprini:2017ins}.

We start with summarizing the analytic structure of a generic hadronic form factor $F(q^2)$ for two pseudoscalar mesons $P_1$ and $P_2$ with
masses $M_1$ and $M_2$, respectively, with $M_1 \geq M_2$; see Ref.~\cite{Caprini:2019osi} for a textbook reference.
This type of form factor emerges in the description of the semileptonic decay $P_1 \to P_2 \bar{\ell} \ell'$,
and further processes related to this decay by crossing symmetry.
In fact, the form factor describes the dynamics of the interaction of these mesons in three distinct regions
of phase space \emph{simultaneously}.
First, it describes the scattering process $P_1 \ell \to P_2 \ell'$ for $q^2 \leq 0$.
Second, it describes the semileptonic decay $P_1 \to P_2 \bar{\ell} \ell'$ for $(m_\ell + m_{\ell'})^2 \leq q^2 \leq (M_1 - M_2)^2 \equiv t_-$.
Third, it describes the pair production process $\ell \bar{\ell}' \to \bar{P}_1 P_2$ for $q^2 \geq t_+ \equiv (M_1 + M_2)^2$.
Here, we only consider a single form factor that is free of singularities to the left of $t_-$.
We note at this point already that in this case, the overall phase of $F$ can be chosen such that $\arg F(q^2) = 0$ for $q^2 \leq t_-$.
Hence, the form factor fulfils the Schwarz reflection condition, \ie, $F(q^2 + i \epsilon) = F^*(q^2 - i\epsilon)$ for $q^2 \geq t_+$
and $\operatorname{Disc}_{q^2} F(q^2)$ is purely imaginary.

When parametrising the form factor for the scattering and decay regions of the phase space,
one commonly uses a dispersively-bounded parametrisation of the form factor $F(q^2)$~\cite{Okubo:1971jf,Boyd:1994tt,Caprini:1997mu}:
\begin{equation}
    \label{eq:prelim:common-param}
    F(q^2) = \frac{1}{\phi_F(q^2) P_\text{sub}(z(q^2))} \sum_{n=0}^{\infty} a_n z(q^2)^n\,.
\end{equation}
Here $z(q^2)$ represents a conformal mapping of the complex $q^2$ plane to the open unit disk $|z| < 1$,
\begin{equation}
    \label{eq:conformal-mapping}
    z(q^2; t_+, t_0) = \frac{\sqrt{t_+ - q^2} - \sqrt{t_+ - t_0}}{\sqrt{t_+ - q^2} + \sqrt{t_+ - t_0}}\,,
\end{equation}
where $t_0$ is an arbitrary parameter $-\infty < t_0 < t_+$ that can be chosen to improve the convergence of the sum in the parametrisation.
Moreover, $\phi_F$ represents the so-called outer function, which is fully determined in the construction of the dispersive bound,
and $P_\text{sub}$ represents the product of so-called Blaschke factors accounting for all below-threshold poles, \ie, poles due to the
presence of bound states with mass $M$ where $t_- < M^2 < t_+$.
The usage of Blaschke factors to account for the below-threshold poles ensures that $|P_\text{sub}|=1$ for $t_+ \leq q^2$.In the following, we assume the absence of below-threshold poles
to streamline the notation --- including such poles is a straightforward exercise.
As a consequence, we use $P_\text{sub} = 1$ from this point forward.

This form of parametrisation benefits from the presence of a dispersive bound
\begin{multline}
    \label{eq:prelim:bound}
    \int_{t_+}^\infty dq^2 \, \omega_F(q^2) |F(q^2)|^2 = \frac{1}{2\pi} \oint \frac{dz}{i z} \left|\sum_{n=0}^\infty a_n z^n\right|^2\\
    = \sum_{n=0}^\infty |a_n|^2 < 1\,.
\end{multline}
Here $\omega_F(q^2)$ is determined in the derivation of the bound and positive definite on the support of the integral.
Consequently, $\phi_F$ is chosen to compensate for the presence of $\omega_F$ and the Jacobian of the change of variable from $q^2$ to $z$.

We show here that a parametrisation of the type shown in \cref{eq:prelim:common-param} can be extended into the pair
production region of phase space ($q^2 > t_+$) with a few modifications.
Our new parametrisation ansatz takes the form
\begin{equation}
    \label{eq:prelim:new-param}
    F(q^2) = \frac{W(z)}{\phi_F(z)}
    \frac{f(z)}{[z - z_r]\,[z - z_r^*]}\bigg|_{z=z(q^2)}\,,
\end{equation}
\text{with}
\begin{equation}
    f(z) = \sum_{n=0}^{N} b_n z^n\,.
\end{equation}
The above differs from the common prescription in two aspects.
\begin{itemize}
    \item First, a weight function $W(z)$ is available to counteract the pathological behaviour of the outer function $\phi_F$
    in the timelike region at $z \to \pm 1$. 
    This type of pathological behaviour has been previously discussed in Ref.~\cite{Buck:1998kp}
    in the context of the timelike pion form factor and in Refs.~\cite{Becher:2005bg,Bourrely:2008za} in the context of the
    $B\to \pi$ vector form factor.
    We further use $W$ to manifestly ensure the correct asymptotic behaviour in the limit $z \to 1$ (\eg, $F (q^2 \to \infty) \sim 1/q^2$).
    \item Second, the factor $[z - z_r]\,[z - z_r^*]$ in the denominator of \cref{eq:prelim:new-param} accounts for the presence of
    an above-threshold resonance $r$ in the pair production process $\ell \bar{\ell}' \to r \to \bar{P}_1 P_2$,
    with mass $M_r$ and decay width $\Gamma_r$.
    This resonance gives rise to two complex-conjugate poles of the form factor $F$ at $z_r$ and $z^*_r$
    on its 2nd Riemann sheet~\cite{Caprini:2017ins}. The conformal mapping \cref{eq:conformal-mapping} maps the 2nd Riemann sheet
    onto the outside of the unit circle in $z$, yielding
    \begin{equation}
        1 / z_r = z\left( \left(M_r - i \Gamma_r/ 2 \right)^2 \right) \,.
    \end{equation}
    The above pole position fulfils $|z_r| > 1$ and $\Im z_r < 0$.
    The presence of two complex-conjugate poles is necessary to ensure that $F$ fulfils the Schwarz reflection
    principle, \ie, that $F$ is real-valued for $q^2 \in \mathbb{R}$ with $q^2 < t_+$~\cite{Caprini:2017ins}.
    The generalisation of our ansatz to the case of multiple relevant above-threshold resonances is straightforward.
\end{itemize}

Despite our best efforts, we are currently not able to maintain the manifest orthogonality of the contributions
to the saturation, \ie, the bound is not manifestly a sum of positive definite quantities.

We emphasise that the factor $[z - z_r]\,[z - z_r^*]$ introduced in \cref{eq:prelim:new-param}
cannot --- in general --- be replaced by a product of Blaschke factors
for our purpose. The reason is that a Blaschke factor describing a pole on the 2nd Riemann sheet
is always accompanied by a zero on the 1st Riemann sheet.
However, a variety of form factors are known to exclude the presence of zeros
on the first Riemann sheet; see, \eg, Refs.~\cite{Leutwyler:2002hm,Ananthanarayan:2011xt} for the pion vector form factor.
Hence, to leave sufficient flexibility in the parametrisation to avoid these additional zeros,
we choose not to use Blaschke factors.

The convergence of a series expansion as shown in \cref{eq:prelim:new-param} is discussed in Ref.~\cite{Buck:1998kp}.
Due to the branch cut in $q^2$, $F$ is defined in the timelike region for momentum values with infinitesimal imaginary parts.
The conformal mapping \cref{eq:conformal-mapping} maps these values to the inside of the unit disk, \ie, in the region of convergence of the series expansion.
Abel's theorem further ensures that the series converges towards its value on the unit circle when it exists.

\section{Application to the Pion Form Factor}
\label{sec:results}

As a proof-of-concept analysis, we study the application of our parametrisation to $\Fpi$. In the isospin symmetry limit,
this form factor corresponds to the isospin $I=1$ projection of the electromagnetic pion form factor $F_\pi^\text{e.m.}$.
Limiting the analysis to $\Fpi$ avoids the complexity of $\rho$--$\omega$ mixing~\cite{Holz:2022hwz,Colangelo:2022prz} and ensures the absence of poles due to $I=0$ resonances.
The form factor $\Fpi$ is defined by
\begin{equation}
    (p_1 - p_2)^\mu \Fpi(q^2) = \frac{1}{\sqrt{2}} \bra{\pi^-(p_1) \pi^0(p_2)} \bar{d}\gamma^\mu u \ket{0}\,,
\end{equation}
where $q^2 = (p_1 + p_2)^2$. Our choice of this particular form factor is motivated by the fact that
it is free of below-threshold poles and that ample experimental data is available
to determine $|\Fpi|^2$ both at timelike and at spacelike $q^2$.
In this proof-of-concept study, we set out to describe the pion form factor up to $\Re q^2 \leq 1\,\GeV^2$.
In this region, the $I=1$ form factor has a single resonance above threshold, which corresponds to the $\rho(770)$ meson.
We determine the pole mass and width of the $\rho$ as part of our fit.
Analyticity-based descriptions of the pion vector form factor (see, \eg, Refs.~\cite{%
Caprini:1999ws,DeTroconiz:2001rip,Leutwyler:2002hm,Colangelo:2003yw,deTroconiz:2004yzs,Hanhart:2012wi,%
Ananthanarayan:2013zua,Ananthanarayan:2016mns,Hanhart:2016pcd,Ananthanarayan:2018nyx%
}) are often based on the Omn\`es representation~\cite{Omnes:1958hv},
which describes the elastic part of $\Fpi$ in terms of the pion--pion $P$-wave phase shift $\delta_1^1$,
\begin{equation}
    \label{eq:results:omnes}
    \Omega(q^2) = \exp \bigg( \frac{q^2}{\pi}\int_{4M_\pi^2}^\infty dt \frac{\delta_1^1(t)}{t(t-q^2)} \bigg)\,.
\end{equation}
We do not use the phase as input here but rather check \textit{a-posteriori} its consistency with studies of Roy equations~\cite{%
Ananthanarayan:2000ht,Colangelo:2001df,Garcia-Martin:2011iqs}, after fitting data on the form factor modulus.
For dispersion relations reconstructing the form factor phase from its modulus, see Ref.~\cite{RuizArriola:2024gwb} and references therein.
By construction, our parametrisation~\cref{eq:prelim:new-param} fulfils~\cref{eq:results:omnes} for $\Omega = |\Fpi|$ and $\delta_1^1 = \arg \Fpi$
for every choice of the truncation order $N$.

The outer function as shown in \cref{eq:outer:z} exhibits zeros at $z\to \pm 1$,
implying that $\Fpi \sim 1/\phi_F$ diverges at $z\to \pm 1$. 
However, $\Fpi$ is finite in both points.
As discussed in Ref.~\cite{Buck:1998kp}, this implies that the weight function $W$ should be chosen
to manifestly remove these spurious divergences rather than letting the series expansion
achieve this as part of the fit.
Moreover, in the limit $q^2 \to \infty$, QCD imposes that $\Fpi(q^2) \sim 1/q^2$~\cite{Lepage:1979zb,Farrar:1979aw} (up to logarithmic corrections),
or equivalently $\Fpi(q^2(z)) \sim (1 - z)^2$.
We choose to manifestly impose this behaviour as part of the weight factor $W$.
Hence, we partially follow Ref.~\cite{Buck:1998kp} in using
\begin{equation}
    W(z) = (1 + z)^{2} (1 - z)^{5/2} \,.
\end{equation}

Armed with a model for the form factor $\Fpi$, we confront it with experimental data
by performing a Bayesian fit using the \EOS software~\cite{EOSAuthors:2021xpv} in version 1.0.13~\cite{EOS:v1.0.13}. Our choice of prior is defined as the product of independent
uniform PDFs. For our nominal fit with truncation $N=$, we use the following prior intervals:
\begin{equation}
\begin{aligned}
     0.757\,\GeV & \leq M_\rho      \leq  0.763\,\GeV\,, \\
     0.141\,\GeV & \leq \Gamma_\rho \leq  0.150\,\GeV\,, \\
    -0.08        & \leq b_2         \leq +0.08\,,       \\
    -0.09        & \leq b_3         \leq +0.09\,,       \\
    -0.05        & \leq b_4         \leq +0.07\,,       \\
    -0.02        & \leq b_5         \leq +0.02\,.
\end{aligned}
\end{equation}
This choice of prior contains more than $99\%$ of the marginal posterior.
Further fits are documented as part of the ancillary material~\cite{EOS-DATA-2024-02}.
We emphasise that the coefficients $b_0$ and $b_1$ are fixed by imposing $F(q^2 = 0) = 1$ (due to charge conservation)
and $\Im F(q^2 = t_+) \sim (q^2 - t_+)^{3/2}$ (due to angular momentum conservation).
The latter is achieved by eliminating the term $\Im F(q^2 = t_+) \supset (q^2 - t_+)^{1/2}$;
see Ref.~\cite{Bourrely:2008za} for a similar application for the $B\to \pi$ form factor.
Hence, a truncation at order $N=1$ corresponds to only two fit parameters: the mass $M_\rho$ and the total decay width $\Gamma_\rho$
of the $\rho$ meson.

\begin{figure}[t]
    \includegraphics[width=.49\textwidth]{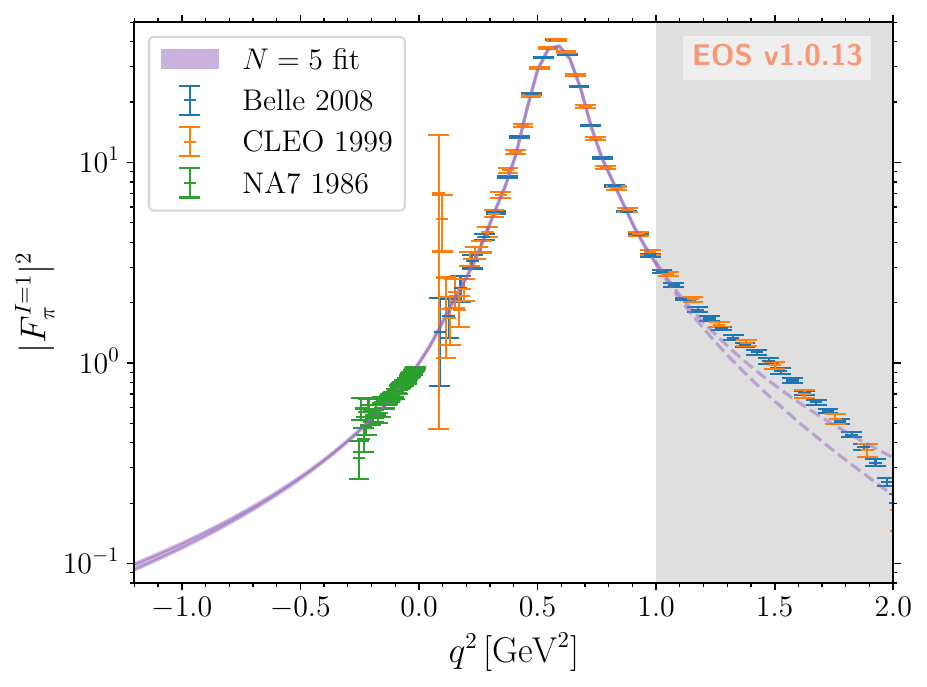}
    \\
    \includegraphics[width=.49\textwidth]{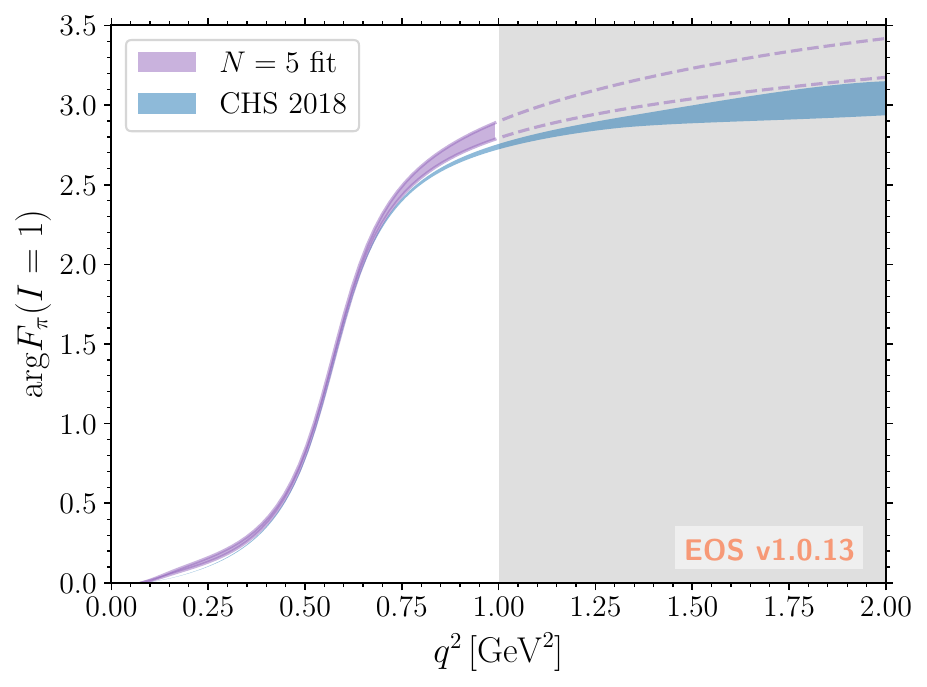}
    \caption{%
        The squared magnitude (top) and the phase (bottom) of the form factor $\Fpi(q^2)$.
        We show the envelope at $68\%$ probability for truncation $N=5$ juxtaposed with data by the Belle, CLEO, and NA7
        experiments. A single data point by the JLab-$F_\pi$ experiment at $q^2 = -2.45\,\GeV^2$
        is not shown but agrees within $0.8\,\sigma$ with the best-fit curve.
        Data points in the shaded region ($q^2 \geq 1\,\GeV^2$) are not fitted.
        However, we extend the posterior-prediction of the $N=5$ fit into
        this region as two dashed curves, indicating only the $68\%$ envelope.
        For reference, we overlay our phase predictions with the input from Ref.~\cite{Colangelo:2018mtw}.
    }
    \label{fig:results:comparison-to-data}
\end{figure}

\begin{table*}[t]
\renewcommand{\arraystretch}{1.2}
\centering
\begin{tabular}{c @{\hskip 3em} ccc @{\hskip 2em} c c c c}
    \toprule
    truncation $N$ & $\chi^2$       & d.o.f.  & $p$ value [\%] & $\langle r_\pi^2\rangle$ [fm$^2$]
                                                                                     & $M_\rho$ [\MeV] & $\Gamma_\rho$ [\MeV] & bound saturation \\
    \midrule
    1              & $\approx 3300$ & $92$    & $< 10^{-10}$   &  ---                & ---             & ---                  & $0.46$ \\
    2              & $\approx 1500$ & $91$    & $< 10^{-10}$   &  ---                & ---             & ---                  & $0.44$ \\
    3              & $117.4$        & $90$    & $2.8$          & $0.474 \pm 0.0022$  & $760.4 \pm 0.4$ & $143.0 \pm 0.6$      & $0.46$ \\
    4              & $98.17$        & $89$    & $23.8$         & $0.457 \pm 0.0045$  & $760.2 \pm 0.4$ & $145.9 \pm 0.9$      & $0.46$ \\
    5              & $97.9$         & $88$    & $22.1$         & $0.460 \pm 0.0061$  & $760.0 \pm 0.6$ & $146.1 \pm 0.9$      & $0.46$ \\

    \bottomrule
\end{tabular}
\renewcommand{\arraystretch}{1.0}
\caption{%
    Goodness-of-fit diagnostics for the fits with truncation $N=1$ through $N=5$
    along-side postdictions of pion charge radius $r_\pi^2$ and the $\rho$
    pole mass \& decay width as well as saturation of the dispersive bound.
}
\label{tab:results:gof}
\end{table*}

The likelihood is comprised of the product of individual, independent contributions:
\begin{description}
    \item[Belle] The Belle experiment has determined $|\Fpi|^2$ at timelike $q^2$ from
    measurements of the decay $\tau^- \to \pi^- \pi^0 \nu$~\cite{Belle:2008xpe}.
    Here, we use the correlated measurements as a multivariate Gaussian likelihood,
    corresponding to $19$ observations.

    \item[CLEO] The CLEO experiment has determined $|\Fpi|^2$ at timelike $q^2$ from measurements
    of the decay $\tau^- \to \pi^-\pi^0 \nu$~\cite{CLEO:1999dln}.
    Here, we use the correlated measurements as a multivariate Gaussian likelihood,
    corresponding to $29$ observations.
    
    \item[NA7] The NA7 experiment has used $\pi H$ scattering to determine
    $|F_\pi^{\text{e.m.}}|^2$ at spacelike $q^2$~\cite{NA7:1986vav}.\footnote{%
\label{note:FpiEM}%
The NA7 experiment extracts from their data the electromagnetic form factor $F_\pi^{\text{e.m.}}$,
which differs from the $I=1$ projection by an isospin-symmetry-violating term $F_\pi^{I=0}$.
For this work, the latter is assumed to be small in the spacelike region.
A simultaneous analysis of both isospin projections is left for future work.
    }
    Here, we assume the data points to be uncorrelated in their statistical error and fully positively
    correlated in their systematical error. We use the NA7 results at spacelike $q^2$
    as a multivariate Gaussian likelihood, corresponding to $45$ observations.

    \item[JLab-$\boldsymbol{F_\pi}$] The JLab-$F_\pi$ experiment
    has used pion electroproduction to determine $F_\pi^\text{e.m.}$ at spacelike $q^2 = -2.45\,\GeV^2$,
    amongst others~\cite{JeffersonLabFpi-2:2006ysh,JeffersonLab:2008jve} (see \cref{note:FpiEM}).
    We use this single data point due to its large leverage,
    \ie, the largest magnitude in $q^2$ available to use experimentally.
    Other data points at $q^2 = -1.6\,\GeV^2$~\cite{JeffersonLabFpi-2:2006ysh}
    and in the interval $q^2 \in [-2.45, -0.6]\,\GeV^2$~\cite{JeffersonLab:2008jve}
    are also obtained;
    however, the correlation between the various data points is not clear to us.
    Since the 2nd data point is found to be well compatible with our nominal fit,
    we conservatively use the $q^2 = -2.45\,\GeV^2$ data point only, corresponding to $1$ observation.

\end{description}
Therefore our fits feature in total $94 - (N + 1)$ degrees of freedom.
We consider any given truncation order to describe the available data well if the corresponding $p$ value exceeds $3\%$.

We carry out a series of fits for truncation orders $N=1$ through $N=5$.
The quality of these fits is evidenced by \cref{tab:results:gof},
together with the predicted pion radius, the $\rho$ resonance parameters,
and the saturations of the dispersive bound that we computed numerically from the left-hand-side integral in \cref{eq:prelim:bound}.

We find that already for $N=1$, the parametrisation visually describes the salient features of the available data.
For the fits with $N=1$ to $N=3$, we find $p$ values below our a-priori threshold.
For $N=4$, we obtain a $p$ value of $\sim 20\%$, 
indicating that the parametrisation starts to describe the data sufficiently well.
Switching from $N=4$ to $N=5$, the $p$ value does not increase, leading us to adopt the $N=5$ fit as our nominal fit,
which is shown in \cref{fig:results:comparison-to-data}.
We further find that the $N=5$ best-fit point is compatible with the $N=4$ best-fit point,
with the additional parameter $b_5^{N=5}$ compatible with zero within its uncertainties.

We note in passing that a fit to only the data in the timelike region yields an excellent description of that data
but suffers from a multimodal posterior, with the individual modes corresponding to solutions with a different number
of zeros on the first Riemann sheet. Using the data in the spacelike region eliminates
all but one of the modes, which corresponds to the absence of any zeros of $\Fpi$ on the 1st Riemann sheet.

Our nominal predictions for the $\rho$ resonance parameters show perfect agreement
and smaller uncertainties than the previous pole determinations~\cite{Garcia-Martin:2011nna,Pelaez:2004xp,Colangelo:2001df},
which are summarised as $M_\rho \in [761, 765]$ MeV and $\Gamma_\rho \in [142, 148]$ in the world average~\cite{PDG2024}.
Our nominal result for the pion charge radius $\langle r_\pi^2\rangle = 0.460(6)~\mathrm{fm}^2$
shows a puzzling tension with the results in the literature~\cite{%
PDG2024,Ananthanarayan:2017efc,Colangelo:2018mtw%
}.
In addition, we find a systematic deviation between the argument of the form factor determined from our fit
and the determinations in the literature~\cite{Colangelo:2018mtw}. These differences motivate a more systematic study of our approach
with respect to the potential impact of further poles and inelastic contributions, which is left to future work.

We find the saturation of the dispersive bound to be in good qualitative agreement with
previous results on the saturation~\cite{Ananthanarayan:2014pta}.

\section{Conclusion}

We have presented a simple parametrisation of hadronic form factors with above-threshold poles.
Crucially, our proposed parametrisation does not require the $\pi\pi$ scattering phase shift as an input,
unlike Omnès-based parametrisations.
We use our proposed parametrisation to fit the isospin-one projection of the pion vector form factor as a first application.
Already at low truncation order $N=4$, \ie, using $M_\rho$, $\Gamma_\rho$ and only three independent shape parameters,
we find an excellent description of the experimental data for $q^2 \leq 1\,\GeV^2$.
We provide posterior predictions of the phase of the form factor as an auxiliary result.
We look forward to promising future applications, such as a simultaneous analysis
of the isospin-one and isospin-zero projections of the pion form factor including
the $\omega$ resonance, or an extension of our current work to higher resonances.
Due to the universality of the final-state interactions, we also envisage the
wider application of our approach to hadronic form factors for flavour-changing currents,
such as in $\bar{B}\to \pi\pi\ell^-\bar\nu$ transitions.

\section*{Acknowledgements}

We are grateful to Florian Herren for valuable discussions and comments on the manuscript.
DvD is grateful to Martin Jung for valuable discussions on a precursor to this analysis.
BK and DvD gratefully acknowledge past support by the DFG through the funds provided to the Sino–-German Collaborative Research Center TRR110
``Symmetries and the Emergence of Structure in QCD'' (DFG Project-ID 196253076 – TRR 110).
DvD~acknowledges ongoing support by the UK Science and Technology Facilities Council (grant numbers ST/V003941/1 and ST/X003167/1).

\appendix

\section{\texorpdfstring{Relation to the $\boldsymbol{\rho}$ Coupling Constants}{Relation to the rho coupling constants}}

Close to the $\rho$ resonance, the $\rho\pi\pi$ and $\rho\gamma$ interactions are described by the Lagrangian~\cite{Hoferichter:2017ftn}
\begin{equation}
    \mathcal{L}_\rho \supset g_{\rho\pi\pi} \epsilon^{abc} \pi^a \partial^\mu \pi^b \rho_\mu^c
        - \frac{e M_\rho^2}{g_{\rho\gamma}} A^\mu \rho_\mu \,.
\end{equation}
In the vicinity of the pole $s_\rho =  (M_\rho - i \Gamma_\rho / 2)^2$, the form-factor $F_\pi$ can be related to these
couplings through its residue on the second Riemann sheet~\cite{Caprini:2017ins, Hoferichter:2017ftn}
\begin{equation}
    \mathrm{Res}(F_{\pi,\mathrm{II}}, s_\rho) = - \frac{s_\rho g_{\rho\pi\pi}}{g_{\rho\gamma}}
\end{equation}
Analyticity and elastic unitarity imply further that the value on the first Riemann sheet is related to the $\rho$ couplings
as well~\cite{Caprini:2017ins, Hoferichter:2017ftn}
\begin{equation}
    F_{\pi, \mathrm{I}}(s_\rho) = -\frac{24 i \pi}{g_{\rho\pi\pi}g_{\rho\gamma}} \left( 1 - \frac{4 M_\pi^2}{s_\rho} \right)^{-3/2}\,.
\end{equation}
As a consequence, our simple analysis of the pion form factors provides full access to both couplings simultaneously.
This is a great advantage over the more involved methods based on the Roy equations, see, \eg, Ref.~\cite{Hoferichter:2023mgy}.
From our nominal fit model, we determine
\begin{equation}
\label{eq:couplings:nominal-results}
\begin{aligned}
    g_{\rho\pi\pi}
        & = 6.09(6) -0.3(1)i = 6.10(6) e^{-0.05(2) i} \\
    g_{\rho\gamma}
        & =  4.87(1) - 0.17(2)i = 4.87(2) e^{-0.035(4) i} \,,
\end{aligned}
\end{equation}
in good agreement with the literature.

\section{Outer Function}
\label{app:outer}

The outer function for the pion vector form factor reads~\cite{Buck:1998kp}
\begin{multline}
    \phi_F(q^2; t_0)
        = \sqrt{\frac{1}{48 \pi \chi}} \left( \frac{t_+ - q^2}{t_+ - t_0} \right)^\frac{1}{4}\\
            \left( \sqrt{t_+ - q^2} + \sqrt{t_+ - t_0} \right) \left( t_+ - t \right)^\frac{3}{4} \\
            \left( \sqrt{t_+ - q^2} + \sqrt{t_+} \right)^{-\frac{1}{2}} \left( \sqrt{t_+ - q^2} + \sqrt{t_+ + Q^2} \right)^{-3} \,,
\end{multline}
where $t_+ = (M_{\pi^+} + M_{\pi^0})^2$ is the threshold, $-\infty < t_0 < t_+$ is a free parameter in the conformal mapping \cref{eq:conformal-mapping},
and $\chi = \chi(Q^2) = 6.839\cdot 10^{-3}\,\GeV^{-2}$ incorporates the normalisation of the dispersive integral that gives rise to a bound of the type
\cref{eq:prelim:bound}. This normalisation is obtained through a perturbative calculation of a suitable
two-point function at a subtraction point $Q^2$~\cite{Buck:1998kp} (see also Ref.~\cite{Simula:2023ujs} for a Lattice QCD estimate).
We choose $Q^2 = - t_0 = 1\,\GeV^2$ for the purpose of this analysis.
The outer function is more conveniently evaluated in terms of $z$ rather than $q^2$:
\begin{multline}
\label{eq:outer:z}
    \phi_F(z)
        = (1+z)^{2} (1-z)^{1/2} \frac{1}{\sqrt{12 \pi t_+ \chi}} \left( 1 - \frac{t_0}{t_+} \right)^{5/4}
    \\
        \left[ \sqrt{1 - \frac{t_0}{t_+}} (1+z) + (1-z) \right]^{-1/2}
    \\
        \times \left[ \sqrt{1 + \frac{Q^2}{t_+}} (1-z) + \sqrt{1 - \frac{t_0}{t_+}} (1+z) \right]^{-3}\,.
\end{multline}

\bibliography{references.bib}

\end{document}